\documentclass[aps,prc,twocolumn,times,graphicx,tighten,showpacs]{revtex4}
\usepackage[dvips]{graphicx}

\newcommand{\bea}{\begin{eqnarray}}
\newcommand{\eea}{\end{eqnarray}}
\newcommand{\be}{\begin{equation}}
\newcommand{\ee}{\end{equation}}

\begin{document}

%%%%%%%%%%%%%%%%%%%%%%%%%%%%%%%%%%%%%%%%%%%%%%%%%%%%%%%%%%%%%%%%%%%%%%

\title{Superscaling Predictions for Neutral Current Quasielastic Neutrino-Nucleus Scattering}

%%%%%%%%%%%%%%%%%%%%%%%%%%%%%%%%%%%%%%%%%%%%%%%%%%%%%%%%%%%%%%%%%%%%%%
\author{M.C. Mart\'{\i}nez$^a$, J.A. Caballero$^b$, T.W. Donnelly$^c$, and J.M. Ud\'{\i}as$^a$}
%%%%%%%%%%%%%%%%%%%%%%%%%%%%%%%%%%%%%%%%%%%%%%%%%%%%%%%%%%%%%%%%%%%%%%

\affiliation{$^a$Grupo de F\'{\i}sica Nuclear, Departamento de F\'{\i}sica At\'omica, Molecular y Nuclear,
Universidad Complutense de Madrid, 28040 Madrid, Spain}
%%%%%%%%%%%%%%%%%%%%%%%%%%%%%%%%%%%%%%%%%%%%%%%%%%%%%%%%%%%%%%%%%%%%%%%
\affiliation{$^b$Departamento de F\'{\i}sica At\'omica, Molecular y Nuclear,
Universidad de Sevilla, 41080 Sevilla, Spain}
%%%%%%%%%%%%%%%%%%%%%%%%%%%%%%%%%%%%%%%%%%%%%%%%%%%%%%%%%%%%%%%%%%%%%%%
\affiliation{$^c$Center for Theoretical Physics, Laboratory for Nuclear
  Science and Department of Physics, Massachusetts Institute of Technology,
  Cambridge, MA 02139, USA}

%%%%%%%%%%%%%%%%%%%%%%%%%%%%%%%%%%%%%%%%%%%%%%%%%%%%%%%%%%%%%%%%%%%%%%%

\begin{abstract}

The application of superscaling ideas to predict
neutral-current (NC) quasielastic (QE) neutrino cross sections is investigated.
 Results obtained within the
relativistic impulse approximation (RIA) using the same relativistic
mean field potential (RMF) for both initial and final nucleons --- a
model that reproduces the experimental $(e,e')$
scaling function --- are used to illustrate the ideas involved.
While NC reactions are
not so well suited for scaling analyses, to a large extent
the RIA-RMF predictions do exhibit superscaling.
Independence of the scaled response on the nuclear species is very
well fulfilled. The RIA-RMF NC superscaling function is in good
agreement with the experimental $(e,e')$ one. The idea that electroweak processes can be described with a universal scaling function, provided
that mild restrictions on the kinematics are assumed, is shown to be valid.

\end{abstract}
%%%%%%%%%%%%%%%%%%%%%%%%%%%%%%%%%%%%%%%%%%%%%%%%%%%%%%%%%%%%%%%%%%%%%%%%
\pacs{25.30.Pt; 13.15.+g; 24.10.Jv}
%%%%%%%%%%%%%%%%%%%%%%%%%%%%%%%%%%%%%%%%%%%%%%%%%%%%%%%%%%%%%%%%%%%%%%
%%%%%%%%%%%%%%%%%%%%%%%%%%%%%%%%%%%%%%%%%%%%%%%%%%%%%%%%%%%%%%%%%%%%%%%%%

\maketitle

Analyses of on-going and future experimental studies of neutrino
reactions and oscillations at intermediate energies~\cite{neut_exp}
inevitably involve nuclear targets and require accurate control of
nuclear effects. One way of taking nuclear effects into account is
by directly modeling them. This approach can predict the bulk of the
neutrino-nucleus response, but is not capable of yielding
predictions of high enough accuracy, given the present experimental
demands. A second approach that has been recently proposed takes
advantage of scaling ideas.
Indeed, scaling has been extensively employed to analyze inclusive
QE electron-nucleus scattering data~\cite{West74,DS199}. The data,
when appropriately organized, scale to a function that is not only
relatively independent of the momentum transfer (scaling of the
first kind), but also independent of the nuclear target (scaling of
the second kind). The simultaneous occurrence of both kinds of
scaling is known as superscaling~\cite{DS199}.
Based on these ideas, a phenomenological SuperScaling Approach
(SuSA)~\cite{Amaro:2006tf,neutrino1} can be pursued that provides a
more robust way to inter-relate the various classes of electroweak
processes than most direct modeling does, as long as the kinematics
chosen lie in the regions where scaling applies, i.e., QE kinematics
for transferred momentum in the range from roughly 500 MeV/c to a
few GeV/c. Within SuSA, one assumes that {\it at similar kinematics}
both electron and neutrino mediated inclusive scattering reactions
share the same universal scaling function, which contains the
relevant information about the initial and final state nuclear
dynamics explored by the probe, thereby allowing one to provide
reliable and relatively model-independent predictions for
neutrino-induced processes employing the $(e,e')$ experimental
scaling function as
input~\cite{Amaro:2006tf,neutrino1,Caballero:2005sj,JA06,Amaro:2006if,
Amaro:2005dn,Caballero:2007tz,Martini:2007jw,Antonov:2006md,Amaro:2006pr}.

To date, most applications of scaling ideas to neutrino-nucleus
cross sections involved charged current (CC) processes, whose
kinematics parallel the electron scattering case. However, the
interaction of neutrinos with matter is mediated not only by $W^\pm$
bosons, but also by the neutral $Z^0$ boson.
NC processes are relevant for oscillation experiments --- for
instance, it is expected that they contribute as the third most
important event type for the MiniBooNE experiment at
Fermilab~\cite{neut_exp}. As in the case of CC processes,
predictions based on scaling ideas, when possible, are clearly
demanded. The identification of CC events is relatively simple via
the outgoing {\it charged} lepton, similar to what happens in
inclusive $(e,e')$ scattering. This means that the energy and
momentum transferred at the leptonic vertex are known and thus the
scaling analysis of CC neutrino-nucleus cross sections proceeds in a
way identical to the electron case. However, in the case of NC
events, the scattered neutrino is not detected and identification of
the NC event is usually made when i) no final charged lepton is
found and ii) a nucleon ejected from the nucleus is detected. Even
in the case that the nucleon energy and momentum can be measured,
the transferred energy and momentum at the leptonic vertex will
remain unknown. The kinematics of the NC process is thus different
from both electron scattering and its CC neutrino counterpart,
rendering the derivation of scaling less obvious. Nevertheless, the
translation of the scaling analysis to NC processes was recently
outlined in \cite{Amaro:2006pr}. There it was shown that the
superscaling analysis of NC reactions in the case of the
Relativistic Fermi Gas (RFG) and scattering of 1 GeV neutrinos from
$^{12}$C is feasible. Said study showed how to extend the scaling
analysis to NC processes. The RFG $(e,e')$ response exhibits perfect
superscaling by definition~\cite{Alberico:1988bv}, but it is not
 in accord with the
magnitude or with the shape of the experimental scaling function. It
has been shown that strong final-state interactions (FSI) are needed
to describe successfully the magnitude and shape of the superscaled
data, introducing also small deviations from the extracted
superscaling behaviour.

In this Letter, we address two crucial questions which arise when
extending SuSA analyses to NC neutrino scattering in the QE region:
i) does superscaling hold for NC neutrino-nucleus cross sections
when strong FSI are present? If so, ii) can the $(e,e')$
experimental scaling function be employed to predict NC cross
sections, in spite of the intrinsic differences between the two
processes? To answer these questions, being aware of how scarce NC
neutrino-nucleus cross section data are, we use predictions from the
Relativistic Impulse Approximation
(RIA)~\cite{Alb97,Chiara03,Martinez:2005xe,JA06,Caballero:2005sj,Jin92},
based on strong relativistic mean field potentials for both the
bound and ejected nucleons (RIA-RMF). This model, as well as its
corresponding semirelativistic version~\cite{Amaro:2006if},
reproduces the shape and magnitude of the experimental scaling curve
extracted from QE $(e,e')$ data, elusive for other theoretical
models. Furthermore, RIA-RMF predicts a universal scaling function
for both electron and CC neutrino
scattering~\cite{JA06,Caballero:2005sj,Caballero:2007tz}.
Here, we verify for the first time that NC QE neutrino cross sections exhibit superscaling
properties even in presence of strong FSI. Insights into the universal character of the
scaling function, {\it i.e.,} the existence of a unique function that simultaneously describes
QE electron, CC and NC neutrino scattering on nuclei, are also provided.

In NC QE neutrino scattering an outgoing nucleon (mass $m_N$) having
energy $E_N$, kinetic energy $T_N = E_N - m_N$ and angle
$\theta_{kp_N}$ with respect to the momentum $\mathbf{k}$ of the
beam is assumed to be detected. The beam energy $\varepsilon$ is
also assumed to be known. These variables determine the kinematics
of the process~\cite{Amaro:2006pr,Barbaro:1996vd}. With regards to the model we employ, the NC QE
neutrino-nucleus scattering is described within the impulse
approximation (IA), where the nuclear current is written as a sum of
single-nucleon currents. The bound nucleon states are given as
self-consistent Dirac-Hartree solutions, derived within a RMF
approach using a Lagrangian containing $\sigma$, $\omega$ and $\rho$
mesons~\cite{boundwf}. FSI effects are included by means of the same
strong RMF potentials that describe the initial bound states. A more
detailed description of the model can be found
in~\cite{Udias,Alb97,Chiara03,Martinez:2005xe}.

\begin{figure}[tph]
{\par\centering
\resizebox*{0.5\textwidth}{0.2\textheight}{\rotatebox{270}
{\includegraphics{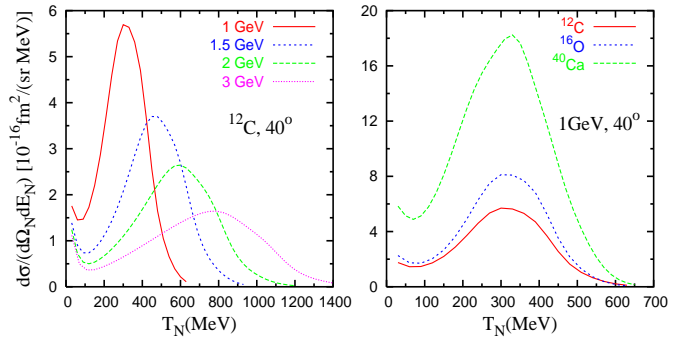}}} \par}
\caption{
  NC QE differential cross section $d\sigma/(dE_N d\Omega_N)$
  versus the outgoing proton kinetic energy $T_N$ for the reaction
  $(\nu,p)$. The left-hand panel corresponds to $^{12}$C at different
  incident neutrino energies $\varepsilon$. The right-hand panel shows
  results at fixed $\varepsilon=1$ GeV for different target nuclei. In
  both panels, the outgoing nucleon detection angle is $\theta_{kp_N}=40^o$.}
\label{Fig1}
\end{figure}

\begin{figure}[tph]
{\par\centering
\resizebox*{0.5\textwidth}{0.2\textheight}{\rotatebox{270}
{\includegraphics{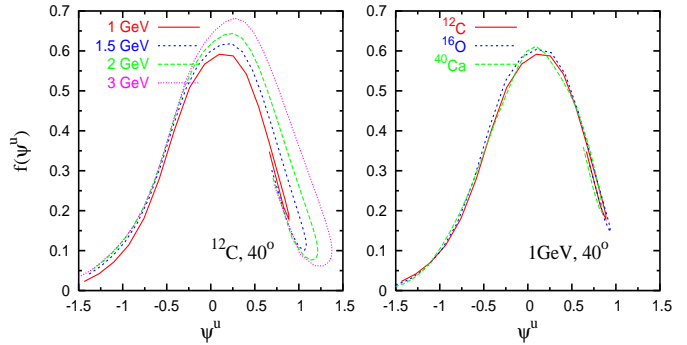}}} \par}
\caption{NC scaling functions corresponding to the differential
  cross sections in Fig.~\ref{Fig1}. }
\label{Fig2}
\end{figure}

 As usual in scaling analyses of QE scattering, we assume the inclusive $A(\nu,N)\nu'X$ cross section
to be obtained as the integrated semi-inclusive one-nucleon (proton or neutron) knockout
$A(\nu,\nu'N)X$ cross sections. In Fig.~\ref{Fig1} we show the strong dependence of NC neutrino
QE inclusive cross sections on the beam energy (provided that
$\theta_{kp_N}$ is fixed), and on the target selected. The results
are obtained with the RIA-RMF model; however a large amount of this
variation is essentially due to the neutrino-nucleus coupling
strength and the variation in the position of the quasielastic peak
for the different beam energies. If superscaling holds, most of this
dependence disappears when dividing these cross sections by the NC
single-nucleon cross section given in Eq.~(20) of
\cite{Amaro:2006pr} and plotting against the dimensionless scaling
variable $\psi^u$ extracted from the RFG analysis in NC kinematics
(see Eq.~(26) in \cite{Amaro:2006pr} for its explicit expression).
The differences in nuclear species should also be taken into
account by the superscaling analysis. Results for the so-obtained
scaling function $f(\psi^u)$ are presented in Fig.~\ref{Fig2}.

In the left-hand panel of Fig.~\ref{Fig2}, one can see
that first-kind scaling is well respected within RIA-RMF in
the region of negative $\psi^u$-values. In other words, the large
variations in the cross sections observed for different neutrino
energies, are accounted for by the single-nucleon part of
the cross sections, which has been factored-out in obtaining the
scaling function. Furthermore, the peak of the
superscaling response appears approximately at the same point for all the kinematics.
However, first-kind scaling is not perfect, as there is a sizeable increase
in the height of the peaks of the curves, as well as a shift to $\psi^u>0$
for increasing beam energy. This is similar to what is observed in RIA-RMF for the
inclusive $(e,e')$ case. Actually, the experimental $(e,e')$ data do
leave room for some breaking of first-kind scaling in the region of
positive scaling variable. First-kind scaling is very well fulfilled for
 electron, CC and NC cases {\it in the absence of
FSI}~\cite{JA06,Caballero:2005sj,Cris07,Alberico:1988bv,Amaro:2006pr}.
Therefore, the breakdown of scaling in Fig.~2 must be ascribed
(within IA) to FSI. In the plane-wave limit, the dependence of the
cross section on the energy of the outgoing nucleon comes mainly
from kinematical effects that are taken into account in the scaling
analysis. However, FSI involve a redistribution of strength that
depends on the energy of the final nucleon. In other words, FSI
introduce an additional, non-kinematical, dependence of the cross
section on $T_N$. If the kinematics of the process are such that the
range of energies of the ejected nucleon depends strongly on the
beam energy, the nucleon will be subject to different FSI for each
$\varepsilon$, and a visible breakdown of first-kind scaling will
show up. This is what happens for $\theta_{kp_N}=40^o$, where there
is a strong shift of the position of the peak of the cross section
with incoming beam energy. However, for those kinematics for which
the range of $T_N$ remains approximately the same when considering
different beam energies, first-kind scaling is obtained even with
FSI included, as FSI effects on the knockout nucleon are similar for
different beam energies.

\begin{figure}[tph]
{\par\centering
\resizebox*{0.35\textwidth}{0.2\textheight}{\rotatebox{270}
{\includegraphics{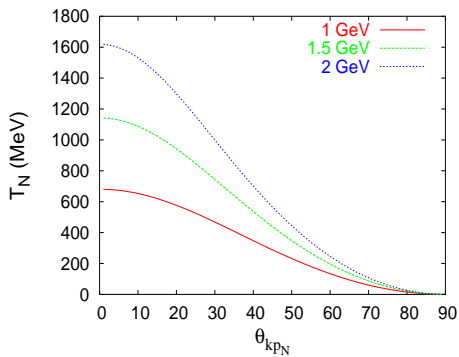}}} \par}
\caption{Relationship
between $\theta_{kp_N}$ and $T_N$ for the NC $(\nu,p)$ reaction on free
protons at rest. The different curves show the results for three
beam energies, 1, 1.5 and 2 GeV.} \label{Fig4}
\end{figure}

Incidentally, in Fig.~2 we also observe that $f(\psi^u)$ is, for
pure kinematical reasons, a bivalued function of the scaling
variable $\psi^u$, as the same value of $\psi^u$ is obtained, at
fixed beam energy and nucleon angle, for two different values of the
outgoing nucleon energy. In the absence of FSI (as in ref.
\cite{Amaro:2006pr}), superscaling is a good approximation and the
two values of the superscaling function for these $\psi^u$ are
nearly equal. When FSI are present, and if the kinematics prevents
superscaling, the bivalued nature of the superscaled function is
revealed.

In order to understand for what kinematics good scaling of the first
kind is reached even in presence of strong FSI, we look at the case
of free nucleons. Fig.~\ref{Fig4} shows how $\theta_{kp_N}$ and
$T_N$ are related due to energy and momentum conservation for
several beam energies. For bound nucleons, neglecting Fermi motion,
the cross section will be peaked at approximately the same $T_N$
value. From the figure one sees that the range of $T_N$ spanned at
fixed $\theta_{kp_N}$ for varying $\varepsilon$ is reduced for large
angles and thus scaling of the first-kind will be much better
obeyed. In general, smaller angles show larger first-kind scaling
violations, while larger angles exhibit almost perfect first-kind
scaling~\cite{Cris07}. Note that this result comes through purely
kinematical reasoning and thus is model-independent to the extent
that the cross section can be described within IA.

Results for scaling of the second kind are presented in the
right-hand panel of Fig.~\ref{Fig2}. The superscaling functions
obtained for several nuclei are almost identical, in spite of the
strong difference in magnitude of the corresponding cross sections
({\it cf.} Fig.~\ref{Fig1}). That is, the dependence on the nuclear
species is well accounted for by the superscaling analysis. Scaling
of second kind is seen to be very robust, thereby opening up a means
of taking into account nuclear effects for different nuclei
employing superscaling ideas.

The superscaling  properties exhibited by NC QE neutrino-nucleus
scattering suggest exploring the validity of the universal character of the scaling function
for inclusive electroweak processes on nuclei, using either
electrons or CC and NC neutrino probes. To the extent that this
universality holds, the phenomenological SuSA approach formerly
applied to predict CC neutrino-nucleus cross sections could also
provide reliable, largely model-independent predictions
for NC processes.

\begin{figure}[tph]
{\par\centering
\resizebox*{0.35\textwidth}{0.2\textheight}{\rotatebox{270}
{\includegraphics{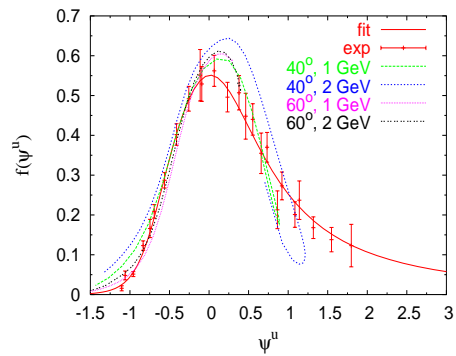}}} \par}
\caption{ NC scaling function evaluated within the RIA-RMF
 approach for 40 and 60 degrees at two beam energies, compared with the
averaged experimental function, together with a phenomenological
parameterization of the data (solid).} \label{Fig3}
\end{figure}

In order to study whether or not this universality assumption holds
also for NC processes, in Fig.~\ref{Fig3} we compare the RIA-RMF NC
superscaling function with the averaged QE experimental function
obtained from the analysis of $(e,e')$ data, together with a
phenomenological parameterization~\cite{DS199,neutrino1,jourdan}.
The RIA-RMF superscaling function has been plotted for two values of
$\theta_{kp_N}$ for which scaling of first-kind is well fulfilled
(60$^o$) or not-so-well (40$^o$). Results are shown for two beam
energies. As seen, the model gives rise to a NC scaling function
that follows closely the behaviour of the $(e,e')$ function and the
bivalued behaviour of the superscaling function is hardly visible
whenever superscaling is well respected (60$^o$). For the case of
40$^o$, breakdown of first-kind scaling is clear, the departure from
the SuSA $(e,e')$ response visible and the bivalued nature of the NC
superscaled function is enhanced. We notice that all curves would
coincide if superscaling was exactly fulfilled in both NC and
$(e,e')$ cases. Since the $(e,e')$ and NC scaling curves are
obtained under rather different kinematical situations, the scaling
curves depart from each other when superscaling is not a good
approximation. This supports the assumption that, under proper
kinematics restrictions, a universal QE scaling function exists
which is valid not only for inclusive electron and CC neutrino
reactions, as seen in~\cite{JA06,Caballero:2005sj}, but also for NC
processes.

In summary, we have established sufficient conditions under which a
universal superscaling function could be applied both to electron and
neutrino (CC or NC) inclusive scattering. These conditions refer to the fact that the
 kinematics  must be such that the range of energies spanned by the ejected nucleon is nearly
independent of the incoming neutrino energy. This happens, for
instance, when the angle of the ejected nucleon with regards to the
beam is larger than roughly $50^o$, which happens to be the region
where the cross section integrated over angles has larger values. In
such a case, first-kind scaling is well respected at the 10$\%$ level
 even in the
presence of strong FSI, and the good comparison with the
experimental $(e,e')$ scaling function gives us confidence that SuSA
can be extended to predict NC QE neutrino cross sections.

We also note that even though  we have illustrated this study within
the RIA-RMF model that contains strong FSI and is quite successful
in reproducing the experimental electron scattering scaling
function, the kinematical conditions that grant the validity of SuSA
are model independent when the IA can be safely applied, that is
under QE kinematics with neutrino beam energies from $\sim 500$ MeV
up to a few GeV.

%%%%%%%%%%%%%%%%%%%%%%%%%%%%%%%%%%%%%%%%%%%%%%%%%%%%%%%%%%%%%%%%%%%%%
\section*{Acknowledgements}
Work partially supported by DGI (Spain) and FEDER funds:
 FIS2005-01105, FPA2006-07393, FPA2006-13807 and FPA2007-62216, by the Junta de
 Andaluc\'{\i}a and the INFN-CICYT, and by the Comunidad de Madrid and UCM (910059 'Grupo de F\'{\i}sica Nuclear' and
PR1/07-14895). It was
also supported in part (TWD) by U.S. Department of Energy
Office of Nuclear Physics under contract No. DE-FG02-94ER40818. M.C.M
acknowledges a 'Juan de la Cierva' contract from MEC.
Computations were performed at the 'High Performance Cluster for Physics'  funded by UCM and FEDER funds.

%%%%%%%%%%%%%%%%%%%%%%%%%%%%%%%%%%%%%%%%%%%%%%%%%%%%%%%%%%%%%%%%%%%%
%%%%%%%%%%%%%%%%%%%%%%%%%%%%%%%%%%%%%%%%%%%%%%%%%%%%%%%%%%%%%%%%%%%

\end{document}